\documentstyle[twocolumn,aps,prl,graphicx]{revtex}

\begin{document}

\draft

\title{Social games in a social network}

\author{Guillermo Abramson$^{1,2}$
\thanks{E-mail address: abramson@cab.cnea.gov.ar}
and Marcelo Kuperman$^1$
\thanks{E-mail address: kuperman@cab.cnea.gov.ar}}
\address{$^1$Centro At{\'o}mico Bariloche and Instituto Balseiro, 8400
S. C. de Bariloche, Argentina \\
$^2$Consejo Nacional de Investigaciones Cient{\'\i}ficas y T{\'e}cnicas,
Argentina}

\maketitle

\begin{abstract}
We study an evolutionary version of the Prisoner's Dilemma game,
played by agents placed in a small-world network. Agents are able
to change their strategy, imitating that of the most successful
neighbor. We observe that different topologies, ranging from
regular lattices to random graphs, produce a variety of emergent
behaviors. This is a contribution towards the study of social
phenomena and transitions governed by the topology of the
community.
\end{abstract}


The search for models that account for the complex behavior of
biological, social and economic systems has been the motivation
of much interdisciplinary work in the last decade \cite{haken}.
In particular, the emergence of altruistic or cooperative
behavior is a favorite problem of game theoretical approaches
\cite{maynard}. In this context, the Prisoner's Dilemma game
\cite{axelrod} has been widely studied in different versions, as
a standard model for the confrontation between cooperative and
selfish behaviors, the later manifested by a defecting attitude,
aspiring to obtain the greatest benefit from the interaction with
another individual. It is usually implemented in zero dimensional
systems, where every player can interact with any other. It has
also been studied on a regular lattice, where a player can
interact with its nearest neighbors in an array \cite{nowak}. In
a regular lattice the concept of a $k$-neighborhood is
straightforward. It is composed of the $k$ nearest individuals to
a given one. However, social situations are rarely well described
by such extreme networks. The topology of social communities is
much better described by what has been called small-world
networks \cite{milgram,watts98}. In the version of small worlds
that we use in this work, the ``regular'' $k$-neighborhood of an
individual is modified by breaking a fraction of its $k$ original
links. An equal amount of new links are created, adding to the
neighborhood a set of individuals randomly selected from the
whole system.

We have studied a simple model of an evolutionary version of the
Prisoner's Dilemma game played in small-world networks. The
Prisoner's Dilemma was chosen as a paradigm of a system capable
to display both cooperative and competitive behaviors
\cite{axelrodbook}. The evolutionary dynamics is implemented by
an imitation behavior. It is important to notice that, contrary
to the {\em iterated} version of the game, our players do not
remember past encounters. The emergence of cooperation is only
due to the spatial organization of the players. This is the main
rationale behind the study of the influence of the topological
aspects of the community. We have explored different topologies,
ranging from a regular lattice to random graphs, going through
the small worlds. We have found surprising collective behaviors
corresponding to the small-world systems, put on evidence by the
enhancement of defection in situations where cooperation is the
norm.

We set up a system of $N$ players arranged at the vertices of a
network (as described below). Each player is a pure strategist,
adopting either a cooperative or a defecting strategy. Each
vertex is connected, on average, to other $2K$ vertices, that we
call its neighbors. The edges connecting two vertices enable the
interaction between players. A round of play consists in the
confrontation of every player with all its neighbors. Each one of
these produces a profit for the player, computed with the payoff
matrix:
\begin{equation}
\begin{array}{c|c|c}
&{\bf C} & {\bf D} \\ \hline
{\bf C} & r & s \\ \hline
{\bf D} & t & p
\end{array}
\end{equation}
Each element in the payoff matrix represents the payoff of a
player performing the strategy at left, when confronting a player
that performs the strategy above. A defector ${\bf D}$ obtains
$t$ ---the temptation to defect---when its opponent is a
cooperator (${\bf C}$), who gets $s$ ---the sucker's payoff. Each
of two cooperators obtains a reward $r$, while each of two
defectors is punished with $p$. The Prisoner's Dilemma is played
with $t>r>p>s$ and $2r>t+s$. In this work we use a simplified
version of the game, assuming that $s=p=0$, $r=1$. All the
interesting features of the game are preserved \cite{nowak}, and
we are left with a single parameter to play with.

Let's represent the players with two-component vectors $x$,
taking the values $(1,0)$ for {\bf C}-strategists and $(0,1)$ for
{\bf D}-strategists. The payoff matrix is:
\begin{equation}  \label{matrix}
{\bf A}= \left(
\begin{array}{cc}
1 & 0 \\
t & 0
\end{array}
\right).
\end{equation}
As a result of the confrontation with its neighbors in a single
round, at time $\tau$, each player collects a payoff
\begin{equation}  \label{payoff}
P_i(\tau ) = \sum_{j\in\Omega_i} x_i {\bf A} x_j^T,
\end{equation}
where $\Omega_i$ is the set of neighbors of element $i$. $P_i$ is
the profit earned by a player in a time step, and it is not
accumulated from round to round.

After this, the players are allowed to inspect the profit
collected by its neighbors in that round, adopting the strategy
of the wealthiest among them for the next round of play. If there
is a draw between more than one neighbor, one of them is chosen
at random to be imitated. If the element under consideration is
itself one of the winners of the round, it keeps its own
strategy. That is, explicitly writing the time dependence of the
strategies:
\[
x_{i}(\tau +1)=\left\{
\begin{array}{ll}
x_{i}(\tau ) & \mbox{if}\;P_{i}(\tau )\geq\max (P\in \Omega _{i}) \\
x_{j}(\tau ) & \mbox{if}\;j\in \Omega
_{i}\;\mbox{and}\;P_{j}(\tau )=\max (P\in \Omega _{i}).
\end{array}
\right.
\]

We have found that a small amount of noise is essential to
prevent the system from falling in a frozen state. After a round
of play, we chose one element at random and flip its strategy.
This is enough to keep the system out of equilibrium and allow
transitions between different states.

As a playground for our system, we have used a family of
small-world networks that depend on a parameter $\epsilon $
\cite{watts98}. We start from a regular, one-dimensional,
periodic lattice of coordination number $2K$. We then run
sequentially through each of the sites, rewiring $K$ of its links
with probability $\epsilon$. Running from 0 to 1, this parameter
changes the wiring properties of the network, ranging from a
completely ordered lattice at $\epsilon =0$, to a random network
at $\epsilon =1$. Intermediate values of $\epsilon $ produce a
continuous spectrum of small-world networks. Double connections
between sites, as well as the connection of a site with itself,
are avoided in the construction of the network. Since we neither
destroy nor create links, the resulting network has an average
coordination number $2K$, equal to the initial one. This method,
however, can produce disconnected graphs, that we have avoided in
our analysis. Note that $\epsilon$ is related to the fraction of
modified regular links.

Two magnitudes characterize the topological properties of the
small-world networks generated by the indicated procedure. One of
them, $L(\epsilon )$, measures the typical separation between any
pair of elements in the network. The other, $C(\epsilon )$,
measures the clustering of an element's neighborhood
\cite{watts98}. Ordered lattices are highly clustered, and have
large $L$. Random graphs have short characteristic length and
small clusterization. In between, small worlds can be
characterized by a high clusterization (like lattices) and short
path lengths (like random networks).

The opposing tendencies of cooperation and defection perform
differently for different payoff tables and different topologies,
through the values of $t$ and $\epsilon$. Disregarding
$\epsilon$, one may qualitatively expect that, for sufficiently
high values of $t$ it would pay to defect while, for low values
of $t$, it would be worth to cooperate. In either of these two
extremes, the system would collapse to a state formed only by
defectors (in the first case) or only by cooperators (in the
second case). For intermediate values of $t$ the system would
settle into a mixed state consisting in cooperators and
defectors. Cooperators would thrive through the formation of
clusters, that can resist the invasion by defectors. The
dependency on the topology of the network appears on top of these
three regimes. From the structure of the payoff matrix one may
conjecture that the high values of $t$ referred to above will be
around $t=2$ (where a defector earns twice as much as a pair of
cooperators). Correspondingly, the low values of $t$ will be
around $t=1$ (where a single cooperator earns more than a
defector).

In the following we show the results of simulations performed in
systems with 1000 elements. The initial strategies are assigned
at random with equal probability. Then several hundred rounds are
played to allow for an asymptotic regime to be achieved. All the
results shown are averages over realizations where both the
networks and the initial conditions are randomly chosen,
excluding all disconnected graphs from our analysis.

\begin{figure}
  \centering
  \resizebox{\columnwidth}{!}{\includegraphics{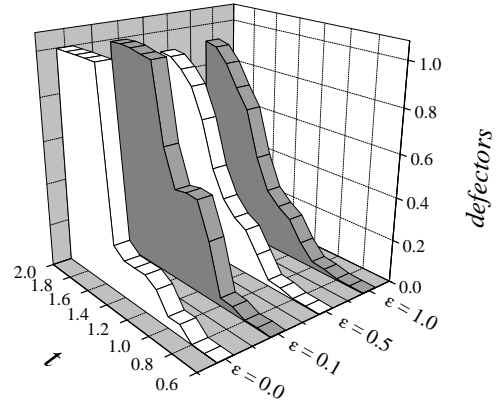}}
  \caption{Fraction of defectors as a function of the temptation
  to defect $t$, for different values of the rewiring probability
  $\epsilon$ (as shown in the legend). The results correspond to
  10 independent realizations
  of 1000 elements, run
  for 500 rounds after a transient of 200 rounds. Note that the range
  of $t$ extends to values lower than 1, where the game is not
  a proper Prisoner's Dilemma.}
  \label{dvsb}
\end{figure}

The number of cooperators and defectors are fluctuating
variables, with bell shaped distributions. In figure \ref{dvsb}
we show the average fraction of defectors in systems with $K=2$,
that is, systems with an average coordination number of four.
Four curves are shown as a function of the parameter $t$. Each
curve corresponds to a  network characterized by the parameter
$\epsilon $ shown in the legend. All the curves show a growth in
the fraction of defectors for growing values of $t$, as expected.
We can see however, that the small world corresponding to
$\epsilon =0.1$ displays an enhanced number of defectors at
values of $t$ around $1.2$. For systems with a fixed $K$ and a
fixed $t$, this means that the existence of a small world
topology with $\epsilon \sim 0.1$ represents that nearly 40\% of
the population adopts the defecting strategy, against the 20\% of
more regular or more random networks. (Note that, in fig.
\ref{dvsb}, we have included values of $t$ lower than 1, where
the game is not a proper Prisoner's Dilemma, since the reward for
cooperation is greater that the temptation to defect. We have
done so because the state of the system at $t=1$, for all values
of $\epsilon$, still contains a small fraction of defectors. We
wanted to stress that for low enough values of $t$ the state is
complete cooperation.)

\begin{figure}
  \centering
  \resizebox{\columnwidth}{!}{\includegraphics{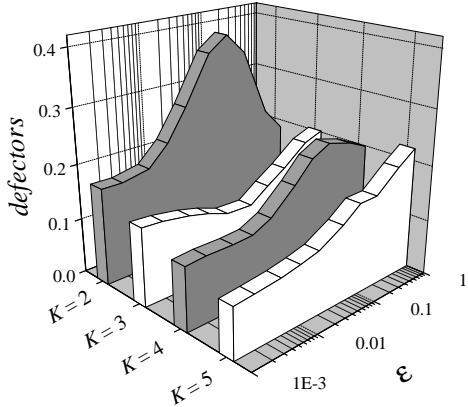}}
  \caption{Fraction of defectors as a function of the rewiring
  probability $\epsilon$, for different values of $K$ (as shown in
  the legend). The results are
  derived from 10 independent realizations of 1000 elements, run
  for 1000 rounds after a transient of 1000 rounds. Games played with
  $t=1.2$}
  \label{dvsp}
\end{figure}

In figure \ref{dvsp} we show a plot of the same system, but with
the fraction of defectors as a function of $\epsilon$, to
emphasize the changes in behavior as the structure of the network
varies. The four curves correspond to different coordination
numbers. The game corresponds to the value $t=1.2$ in the payoff
matrix, so that the curve with $K=2$ is a slice of figure
\ref{dvsb} cut at $t=1.2$. Note that only this curve has a clear
high peak of defectors centered near $\epsilon =0.1$. Systems
with $K=3$ have a downward peak instead, in the region of small
worlds, indicating a slight enhancement of the cooperative
strategy. For $K=4$ we can see again a small peak of defectors.
Systems with $K=5$ and greater (not shown) display a monotonous
behavior in $\epsilon$.

Some conjecture on the origin of these features may be
appropriate here. We think that the competition between the
stability of clusters of cooperators and their exploitation by
neighbor defectors at the borders contributes to the features
observed here. When $K=2$, the cooperators survive in small
compact groups. As $\epsilon$ grows, these groups can be formed
by elements widely dispersed in the system, where they will have
more defector neighbors to compete with. In this way, there will
be less configurations to support them and, consequently, more
defectors in the system. For $\epsilon$ even greater, and more
long range links, cooperators may start to reconnect and survive
the competition with the defectors. When $K>2$, the cooperators
can only survive in larger groups, because defecting neighbors at
the border of a group can penetrate deeper. When $\epsilon$
grows, cooperators belonging to faraway groups may become
connected to form large clusters able to survive. The fact that
$K=3$, at $t=1.2$ shows a slight \emph{decrease} in the fraction
of defectors at intermediate values of $\epsilon$ remains,
however, unexplained in this picture. At other values of $t$, we
observed that the system with $K=3$ performs like that with
$K=2$, namely with an enhancement of defector at intermediate
values of $\epsilon$.

\begin{figure}
  \centering
  \resizebox{\columnwidth}{!}{\includegraphics{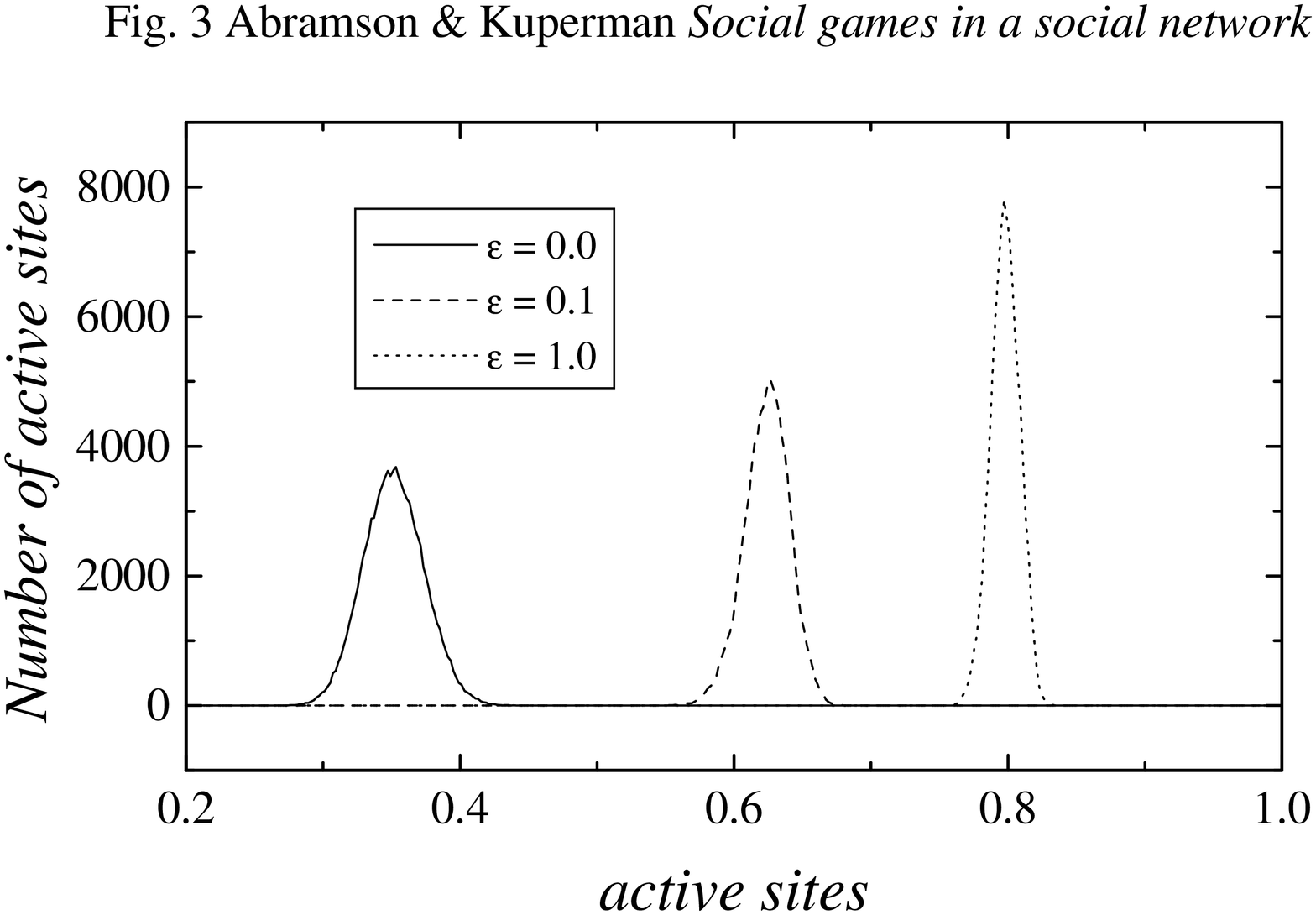}}
  \caption{Distribution of the fraction of ``unsatisfied'' or
  active elements, for different values of the rewiring probability
  $\epsilon$ (as shown in the legend). The results correspond to
  100 independent realizations
  of 1000 elements, run
  for 1000 rounds after a transient of 1000 rounds, with $t=1.2$.}
  \label{unsatisfied}
\end{figure}

Also, we have observed that, for every system, there is a
fluctuating number of ``unsatisfied'' elements that change their
strategy. There is a Gaussian distribution of these unsatisfied
elements, whose mean increases with $\epsilon$, as shown in
figure \ref{unsatisfied}. This behavior is observed for all
values of $K$ and of $t$, namely that regular lattices contain a
smaller number of unsatisfied elements than random networks, with
small worlds in between.

Most of what is analytically known about small worlds refers to
the distribution of shortest paths between pairs of elements (see
for example \cite{newman99,newman00,moukarzel}). It is known that
regular lattices stand apart from even infinitesimally rewired
small worlds, that behave like random networks. The existence of
a phenomenon like the enhancement of defectors density at a
finite value of $\epsilon$, as shown in this work, points to the
existence of an interesting phenomenology in small worlds. The
broad spectrum of behaviors of a given system as a function of the
topological features of the network is the main aspect that we
want to emphasize. This suggests the possibility of modelling a
certain system featuring well known interactions and analyzing
the influence of the particular organization the community.
Moreover, the possibility of a self organizing network with
changing links opens the possibility of modelling more
realistically social and economical situations \cite{zim}. At this
point we can state that the self organization of the network can
lead to a nontrivial behavior of the whole system. Another
interesting example of this statement would be a simple SIR model
for the propagation of an epidemic. This is the subject of work
under way \cite{sir}.

\medskip The authors thank Dami{\'a}n H. Zanette for interesting discussions.

\end{document}